\documentclass[aps,prd,preprint,floats,epsf,superscriptaddress,nofootinbib]{revtex4}

\usepackage{amsmath,amssymb,amsthm,amsfonts,mathtools} 
\usepackage{bm,bbm} 
\usepackage{braket} 
\usepackage{cancel} 
\usepackage{CJKutf8} 
\usepackage{color}
\usepackage{dcolumn} 
\usepackage{dsfont}
\usepackage{graphicx} 
\usepackage{indentfirst}
\usepackage{mathrsfs}
\usepackage{slashed} 
\usepackage{ulem}
\usepackage[utf8]{inputenc} 
\usepackage{xcolor}

\def\ov{\overline}
\def\be{\begin{eqnarray}}
\def\en{\end{eqnarray}}
\def\non{\nonumber}

\def\la{\langle}
\def\ra{\rangle}
\def\A{{\cal A}}
\def\B{{\cal B}}

\def\lsim{ {\ \lower-1.2pt\vbox{\hbox{\rlap{$<$}\lower5pt\vbox{\hbox{$\sim$}
}}}\ } }
\def\gsim{ {\ \lower-1.2pt\vbox{\hbox{\rlap{$>$}\lower5pt\vbox{\hbox{$\sim$}
}}}\ } }

\begin{document}

\font\el=cmbx10 scaled \magstep2{\obeylines\hfill October, 2024}
\vskip 0.8 cm

\title{Tetraquark nature of the $a_0(980)$ meson in hadronic $D$ decays}

\author{Hai-Yang Cheng}
\email[E-mail: ]{phcheng@phys.sinica.edu.tw}
\affiliation{Institute of Physics, Academia Sinica, Taipei, Taiwan 11529, ROC}

\author{Cheng-Wei Chiang}
\email[E-mail: ]{chengwei@phys.ntu.edu.tw}
\affiliation{Department of Physics, National Taiwan University, Taipei, Taiwan 10617, ROC}
\affiliation{Physics Division, National Center for Theoretical Sciences, Taipei, Taiwan 10617, ROC}

\author{Fanrong Xu}
\email[E-mail: ]{fanrongxu@jnu.edu.cn}
\affiliation{Department of Physics, College of Physics $\&$ Optoelectronic Engineering, Jinan University, Guangzhou 510632, P.R. China}

\begin{abstract}
\vskip 0.5cm
The internal structure of the light scalar meson $a_0(980)$ is explored in the three-body $D$ decays of $D\to a_0(980)P\to P_1P_2P$ through the intermediate state $a_0(980)$, where $P$ denotes a pseudoscalar meson. The quasi-two-body $D\to a_0(980)^+P$ decays are governed by the external $W$-emission diagram in which $a_0(980)^+$ is emitted.
The predicted branching fractions in the $q\bar q$ model of $a_0(980)$ are too small by one to two orders of magnitude compared to experiment as the amplitude is suppressed by the smallness of the $a_0(980)^+$ decay constant, while those for $D^+\to a_0(980)^0 P$ and $D^0\to a_0(980)^{-}P$ are usually too large. These discrepancies can be resolved, provided that $a_0(980)$ is a tetraquark state. In this case, there exist two additional $T$-like topological amplitudes, denoted by $\ov T$ and $\tilde T$ which readily account for the discrepancies.  An important implication of the tetraquark model is that the $D_s^+\to a_0(980)^+\pi^0+a_0(980)^0\pi^+$ decay is not a purely $W$-annihilation process as in the $q\bar q$ model of $a_0(980)$; it receives dominant contributions from $\ov T$ newly noticed in this work. Therefore, measurements of $(D,D_s^+)\to a_0(980)P$ decays lend strong support to the tetraquark picture of $a_0(980)$.
\end{abstract}

\maketitle

\small

\section{Introduction} 
\label{sec:intro}

For a long time, the internal structure of the light scalar mesons $\sigma/f_0(500)$, $f_0(980)$, $\kappa/K_0^*(700)$  and $a_0(980)$, whether they are two-quark $q\bar q$ or tetraquark $q^2\bar q^2$ states, is not well established both experimentally and theoretically.  Recently, BESIII has observed $D\to a_0(980)\pi$ modes in the decays of $D^0\to \pi^+\pi^-\eta$ and $D^+\to\pi^+\pi^0\eta$ and obtained the following branching fractions~\cite{BESIII:Dtoa0pi}:
\begin{align}
\begin{split}
& \B(D^0\to a_0(980)^+\pi^-; a_0(980)^+\to \pi^+\eta)=(0.55\pm0.05\pm0.07)\times 10^{-3},
\\
& \B(D^0\to a_0(980)^-\pi^+; a_0(980)^-\to \pi^-\eta)=(0.07\pm0.02\pm0.01)\times 10^{-3},
 \\
& \B(D^+\to a_0(980)^+\pi^0; a_0(980)^+\to \pi^+\eta)=(0.95\pm0.12\pm0.05)\times 10^{-3},
\\
& \B(D^+\to a_0(980)^0\pi^+; a_0(980)^0\to \pi^0\eta)=(0.37\pm0.10\pm0.04)\times 10^{-3},        
\end{split}
\end{align}
which lead to the ratios:
\begin{align}
\label{eq:r}
\begin{split}
r_{+/-} &\equiv {\B(D^0\to a_0(980)^+\pi^-)\over \B(D^0\to a_0(980)^-\pi^+)}= 7.5^{+2.5}_{-0.8}\pm1.7
~, \\ 
r_{+/0} &\equiv  {\B(D^+\to a_0(980)^+\pi^0)\over \B(D^+\to a_0(980)^0\pi^+)}= 2.6\pm0.6\pm0.3
~.        
\end{split}
\end{align}
Naively, the ratios are expected to be much smaller than unity because the quasi-two-body decays $D\to a_0(980)^+P$, governed by the external $W$-emission diagram with $a_0(980)^+$ being an emitted meson, are suppressed by the smallness of the decay constant of $a_0(980)^+$. 
The predicted branching fractions of $D\to a_0(980)^+P$ in the $q\bar q$ model of light scalar mesons are too small by one to two orders of magnitude compared to the experiment (for details, see Table~\ref{tab:BF3body} below). On the contrary, the calculated rates of  $D^+\to a_0(980)^0 P$ and $D^0\to a_0(980)^{-}P$ are usually too large compared to the data.
For example, realistic calculations in Ref.~\cite{Cheng:DtoSP} indicate that $r_{+/-}=0.045$ which is too small by two orders of magnitude compared to the BESIII measurement. This strongly implies that the $q\bar q$ scenario for $a_0(980)$ fails to explain the data.

In the tetraquark model, we shall see that for the above-mentioned decays there exist two additional $T$-like topological 
diagrams which we call $\tilde{T}$ and $\ov T$ (the corresponding diagrams are shown in Fig.~\ref{fig:DtoSPTopdiag.pdf}).  Unlike the external $W$-emission diagram $T'$ in which $a_0(980)^+$ is the emitted meson, they are not suppressed by the smallness of the decay constant of the charged $a_0(980)$ meson. Although these amplitudes are non-factorizable and thus pose difficulty in perturbative calculations, their information can be extracted from the data.  We shall see that these $T$-like topologies will enable us to resolve the large discrepancy between theory and experiment occurring in the $q\bar q$ model of $a_0(980)$.

In the $D_s^+$ sector, the decay $D_s^+\to a_0^+\pi^0+a_0^0\pi^+$ followed by $a_0(980)^{+,0}\to \pi^{+,0} \eta$ has been observed by BESIII with a branching fraction of $(1.46\pm0.27)\%$~\cite{BESIII:Dstoa0pi}. If $a_0(980)$ is a $q\bar q$ state,  this decay proceeds only through the $W$-annihilation diagram. However, its branching fraction at a percent level is much larger than the other two $W$-annihilation channels $D_s^+\to \omega\pi^+$ and $\rho^0\pi^+$ whose branching fractions are $(1.92\pm0.30)\times 10^{-3}$ and $(1.9\pm1.2)\times 10^{-4}$, respectively~\cite{PDG}. Na{\"i}vely, these data would imply that the annihilation amplitude is mysteriously sizable in the $SP$ sector, $|A/T|_{SP}\sim 1/2$~\cite{Cheng:DtoSP}, in stark contrast to the suppressed $|A/T|_{PP}\sim 0.18$ in the $P\!P$ sector and $|A_V/T_P|_{VP}\sim 0.07$ in the $V\!P$ sector~\cite{Cheng:2024hdo}. This is a puzzle as {\it a priori} there is no reason to expect a large enhancement of the $W$-annihilation amplitude in the $D\to SP$ decays.  We find that if $a_0(980)$ is a tetraquark state instead, $D_s^+\to a_0^+\pi^0+ a_0^0\pi^+$ will receive a contribution from $\ov T$ and the predicted rate agrees with experiment even in the absence of the $W$-annihilation amplitude.  This has an important implication that $D_s^+\to a_0^+\pi^0+a_0^0\pi^+$ is not a purely $W$-annihilation process and thus the aforementioned puzzle is readily resolved.

Very recently, BESIII made an observation of $\Lambda_c^+\to \Lambda a_0(980)^+\to \Lambda \pi^+\eta$ with a branching fraction of $(1.05\pm0.16\pm0.05\pm0.07)\%$~\cite{BESIII:LctoLa0}, which is unexpectedly large in the $q\bar q$ model. The factorizable contribution is rather suppressed due to the tiny $a_0^+$ decay constant. An early model calculation based on factorization and the pole model gave a branching fraction of order $1.9\times 10^{-4}$~\cite{Sharma:2009zze}.  It had been improved by one order of magnitude after including final-state rescattering, yielding the result of $(1.7^{+2.8}_{-1.0}\pm0.3)\times 10^{-3}$~\cite{Yu:2020vlt}. In the tetraquark model of $a_0(980)$, however, this decay also proceeds through the diagram $\tilde{T}$ to be introduced below, which allows to reach the branching fraction at a percent level.

The layout of the present paper is as follows.  In Section~\ref{sec:properties}, we discuss the two-quark $q\bar q$ and tetraquark pictures of the scalar nonet near or below 1~GeV and the discrimination of the two scenarios in semileptonic decays of charmed mesons. We turn to hadronic $D\to a_0(980)P$ decays in Section~\ref{sec:hadronic} and emphasize the new topological diagrams unique to the four-quark nature of light scalars. We give the explicit expressions for the factorizable amplitudes and extract the new topologies from the available data. Section~\ref{sec:Conclusions} comes to our conclusions. A detailed calculation of the three-body decay rate is presented in the Appendix. The tetraquark nature of $a_0(980)$ has also been discussed in the early literature; see, e.g., Ref.~\cite{Achasov:2021dvt}.

\section{Physical properties of light scalar mesons}
\label{sec:properties}

The underlying structure of light scalar mesons is not well established even until today. Scalar mesons with masses below or close to 1~GeV include the isoscalars $\sigma/f_0(500)$, $f_0(980)$, the isodoublet $\kappa/K_0^*(700)$ and the isovector $a_0(980)$.  If these scalar meson states are identified as the conventional low-lying $0^+$ $q\bar q$ nonet,
their flavor wave functions read
 \be
 && \sigma={1\over \sqrt{2}}(u\bar u+d\bar d) ~, \qquad\qquad~
 f_0= s\bar s ~, \non \\
 && a_0^0={1\over\sqrt{2}}(u\bar u-d\bar d) ~, \qquad\qquad a_0^+=u\bar d ~,
 \qquad a_0^-=d\bar u ~,  \\
 && \kappa^{+}=u\bar s ~, \qquad \kappa^{0}= d\bar s ~, \qquad~
 \bar \kappa^{0}=s\bar d ~,\qquad~ \kappa^{-}=s\bar u ~, \non
 \en
where an ideal mixing for $f_0$ and $\sigma$ is assumed here as $f_0(980)$ is the heaviest one and $\sigma$ the lightest one in the light scalar nonet.  However, this simple picture encounters several serious difficulties. Two of them are (i) 
it is impossible to understand the mass degeneracy between $f_0(980)$ and $a_0(980)$, which is the so-called ``inverted spectrum problem,'' and (ii) it is difficult to explain why $\sigma$ and $\kappa$ are much broader than $f_0(980)$ and $a_0(980)$ in width.  It turns out that these issues can be readily resolved in the tetraquark scenario where the four-quark flavor wave functions of light scalar mesons are symbolically given by~\cite{Jaffe}
 \be \label{4quarkw.f.}
 && \sigma=u\bar u d\bar d ~, \qquad\qquad\qquad~~
 f_0= \frac{1}{\sqrt2} (u\bar u+d\bar d) s\bar s ~,  \non  \\
 && a_0^0= \frac{1}{\sqrt2} (u\bar u-d\bar d) s\bar s ~,
 \qquad a_0^+=u\bar ds\bar s ~,
 \qquad a_0^-=d\bar us\bar s ~,  \\
 && \kappa^+=u\bar sd\bar d ~, \qquad \kappa^0=d\bar su\bar u ~,
 \qquad \bar \kappa^0=s\bar du\bar u ~,
 \qquad \kappa^-=s\bar ud\bar d ~. \non
 \en
This four-quark description explains naturally the inverted mass spectrum of the light nonet, especially the mass degeneracy between $f_0(980)$ and $a_0(980)$, and accounts for the broad widths of $\sigma$ and $\kappa$ while $f_0(980)$ and $a_0(980)$ are narrow because of the suppressed phase space for their decays to the kaon pairs.

In the na{\"i}ve $q\bar q$ model with ideal mixing for $f_0(980)$ and $\sigma(500)$, $f_0(980)$ is purely an $s\bar s$ state, while $\sigma(500)$ is an $n\bar n$ state with $n\bar n\equiv (\bar uu+\bar dd)/\sqrt{2}$.  However, there also exists some experimental evidence indicating that $f_0(980)$ is not a purely $s\bar s$ state. For example, the observation of $\Gamma(J/\psi\to f_0\omega)\approx {1\over 2}\Gamma(J/\psi\to f_0\phi)$~\cite{PDG} clearly shows the existence of both non-strange and strange quark contents in $f_0(980)$.  Therefore, isoscalars $\sigma(500)$ and $f_0(980)$ must have a nontrivial mixing
\be \label{eq:mixingin2quark}
 |f_0(980)\ra = |s\bar s\ra\cos\theta+|n\bar n\ra\sin\theta ~,
 \qquad |\sigma(500)\ra = -|s\bar s\ra\sin\theta+|n\bar n\ra\cos\theta ~.
\en
Various mixing angle measurements have been discussed in the literature and summarized in Refs.~\cite{CCY,Fleischer:2011au}. A recent measurement of the upper limit on the branching fraction product $\B(\ov B^0\to J/\psi f_0(980))\times\B(f_0(980)\to \pi^+\pi^-)$ by LHCb leads to $|\theta|<30^\circ$~\cite{LHCb:theta}.
Likewise, in the four-quark scenario for light scalar mesons, one can also define a similar $f_0$-$\sigma$ mixing angle
 \be \label{eq:mixingin4quark}
 |f_0(980)\ra =|n\bar ns\bar s\ra\cos\phi
 +|u\bar u d\bar d\ra\sin\phi ~, \qquad
 |\sigma(500)\ra = -|n\bar ns \bar s\ra\sin\phi+|u\bar u d\bar d\ra\cos\phi ~.
 \en
It has been shown that $\phi=174.6^\circ$~\cite{Maiani}.

In principle, the two-quark and four-quark descriptions of the light scalars can be discriminated in the semileptonic charm decays.  For example, the ratio
\be
R={\B(D^+\to f_0(980)\ell^+\nu)+\B(D^+ \to f_0(500)\ell^+\nu)
\over \B(D^+\to a_0(980)^0\ell^+\nu)}
\en
is equal to 1 in the $q\bar q$ scenario and 3 in the four-quark model under the flavor SU(3) symmetry~\cite{Wang:2009azc}. BESIII measurements of $D^+\to a_0(980)^0e^+\nu_e$~\cite{BESIII:Dtoa0SL}, $D^+\to f_0(500) e^+\nu_e$ and the upper limit on $D^+\to f_0(980) e^+\nu_e$~\cite{BESIII:DtosigmaSL} imply that $R>2.7$ at 90\% confidence level and hence favor the SU(3) nonet tetraquark description of the $f_0(500)$, $f_0(980)$ and $a_0(980)$ produced in charmed meson decays.

In the $q\bar q$ model, the strange quark content of $\sigma/f_0(500)$ depends on the $f_0-\sigma$ mixing angle, which is of order $30^\circ$, while in the tetraquark scenario of $\sigma$, its strange content is quite suppressed as the mixing angle defined in Eq. (\ref{eq:mixingin4quark}) is close to $\pi$. As a result, the predicted branching fraction of $D_s^+\to f_0(500)e^+\nu_e\to \pi^+\pi^- e^+\nu_e$ can be very different in the 2-quark and 4-quark models. A recent calculation predicted it to be $(20.3\pm1.8\pm0.5)\times 10^{-4}$ and $(0.58^{+1.43}_{-0.57}\pm0.01)\times 10^{-4}$~\cite{Hsiao:2023qtk}, respectively, while the recent BESIII measurement yielded $\B(D_s^+\to f_0(500)e^+\nu_e\to \pi^+\pi^- e^+\nu_e)<3.3\times 10^{-4}$~\cite{BESIII:2023wgr}.  Evidently, the $q\bar q$ model of $f_0(500)$ is ruled out in this case.

\section{Nonleptonic $D\to a_0(980)P\to P_1P_2P$ decays}
\label{sec:hadronic}

We now turn to the hadronic decays of $D\to a_0(980)P$ followed by $a_0(980)\to P_1P_2$ which can be used to discriminate the internal structure of $a_0(980)$. A suitable framework is the topological diagrammatic approach which we are going to describe below.

\subsection{Topological amplitudes}

\begin{figure}[t]
\begin{center}
\vspace{10pt}
\includegraphics[width=0.8\textwidth]{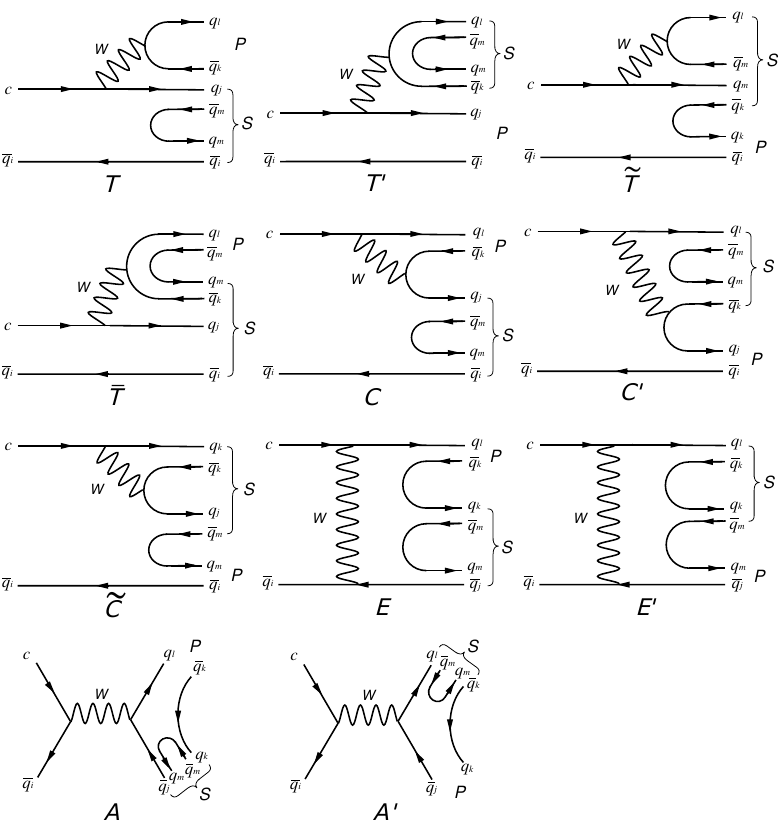}
\caption{Topological diagrams for $D\to SP$ where $S$ stands for a light scalar meson with a tetraquark content.  In the 2-quark model of the light scalars, topological diagrams are the same as that in the tetraquark model except that the diagrams $\ov T$, $\tilde{T}$ and $\tilde{C}$ are absent while the quark pair $q\bar q$ is removed from other diagrams.  
} \label{fig:DtoSPTopdiag.pdf}
\end{center}
\end{figure}

A model-independent analysis of heavy meson decays can be carried out in the so-called topological diagram approach~\cite{Chau,CC86,CC87}. For the purpose of discussing decay rates, it suffices to consider just the tree-type amplitudes: color-allowed tree amplitude $T$, color-suppressed tree amplitude $C$, $W$-exchange amplitude $E$, and $W$-annihilation amplitude $A$. Just as in $D\to V\!P$ decays, one generally has for the $D \to SP$ decays two sets of distinct diagrams for each topology. For example, there are two external $W$-emission and two internal $W$-emission diagrams, depending on whether the spectator quark of the charmed meson stays in the pseudoscalar or scalar meson.   Following the convention in Refs.~\cite{ChengSP,Cheng:SAT}, we shall denote the primed amplitudes $T'$ and $C'$ for the case when the emitted meson is a scalar one (see Fig.~\ref{fig:DtoSPTopdiag.pdf}).  For the $W$-exchange and $W$-annihilation diagrams with the final state $q_1\bar q_2$, the primed amplitude denotes that the even-parity meson contains the $q_1$ quark.

\begin{table}[h!]
\caption{Topological amplitudes of various $D\to a_0(980)P$ decays in the $q\bar q$ and tetraquark models of $a_0(980)$.  Since the topology $\tilde{C}$ is the same as $\tilde{T}$, we will not consider this redundant contribution.
 For simplicity and convenience, we have dropped the mass identification for       
$a_0(980)$. Here $\lambda_{sd}\equiv V_{cs}^*V_{ud}$, $\lambda_{ds}\equiv V_{cd}^*V_{us}$, $\lambda_{d}\equiv V_{cd}^*V_{ud}$ and $\lambda_{s}\equiv V_{cs}^*V_{us}$.
}
 \medskip
\label{tab:DSP}
\begin{ruledtabular}
\begin{tabular}{l l l }
Decay & Quark-antiquark & Tetraquark  \\
 \hline
 $D^+\to a_0^+\ov K^0$ & $\lambda_{sd}(T'+C)$ & $\lambda_{sd}(T'+C+\tilde{T})$   \\
\qquad $\to a_0^+\pi^0$ & ${1\over\sqrt{2}}\lambda_{d}(-T'-C+A-A')$ & ${1\over\sqrt{2}}[\lambda_{d}(-T'-C+A-A')-\lambda_s(\tilde{T}-\ov T)]$ \\
 \qquad $\to a_0^0\pi^+$ & ${1\over\sqrt{2}}\lambda_{d}(-T-C'-A+A')$ & ${1\over\sqrt{2}}[\lambda_{d}(-T-C'-A+A')+\lambda_s(\tilde{T}-\ov T)]$ \\
$D^0\to a_0^+\pi^-$ & $\lambda_{d}(T'+E)$ & $\lambda_{d}(T'+E)+\lambda_s(\tilde{T})$  \\
  \quad~ $\to a_0^-\pi^+$ & $\lambda_{d}(T+E')$ & $\lambda_{d}(T+E')$  \\
  \quad~ $\to a_0^+K^-$ & $ \lambda_{sd}(T'+E)$ & $ \lambda_{sd}(T'+\tilde{T}+E)$  \\
  \quad~ $\to a_0^0\ov K^0$ & ${1\over\sqrt{2}}\lambda_{sd}(C-E)$  & ${1\over\sqrt{2}}\lambda_{sd}(\ov T+C-E)$ \\
  \quad~ $\to a_0^-K^+$ & $ \lambda_{ds}(T+E')$ & $ \lambda_{ds}(T+\ov{T}+E')$  \\
 $D_s^+\to a_0^+\pi^0$ & ${1\over \sqrt{2}}\lambda_{sd}(A-A')$  & ${1\over \sqrt{2}}\lambda_{sd}(A-A'+\ov{T})$  \\
    \quad~ $\to a_0^0\pi^+$ & ${1\over \sqrt{2}}\lambda_{sd}(-A+A')$  & ${1\over \sqrt{2}}\lambda_{sd}(-A+A'-\ov{T})$  \\
\end{tabular}
\end{ruledtabular}
\end{table}

If the light scalar meson is a tetraquark state, there exist additional topological diagrams $\tilde{T}$, $\ov T$ and $\tilde{C}$ as depicted in Fig.~\ref{fig:DtoSPTopdiag.pdf}. It follows that the topology $\tilde{C}$ is equivalent to $\tilde{T}$, and hence this redundant contribution will not be considered in this work.  We shall see below that the diagrams $\tilde{T}$ and $\ov T$ are crucial for understanding the $D\to a_0(980)P$ decays, whose topological amplitude decompositions are listed in Table~\ref{tab:DSP} in both $q\bar q$ and tetraquark models for the light scalar mesons. The topological diagram $\tilde{T}$ has been noticed before in Refs.~\cite{BESIII:DtoKSa0,Achasov:2024nrh}.  However, we find that the new contribution $\ov T$ is comparable to $T$ in size and plays a more dominant role than $\tilde{T}$ in these decays.

\subsection{$D\to a_0(980)P; a_0(980)\to P_1P_2$}

The topological diagrams $T, T', C$ and $C'$ are factorizable, and have the following expressions (see Ref.~\cite{Cheng:DtoSP} for detail):
\begin{align}
\begin{split}
T(a_0\pi) &=-a_1(a_0\pi) f_\pi(m_D^2-m_{a_0}^2)F_0^{D a_0}(m_\pi^2) ~,
\\
T'(\pi a_0) &=a_1(\pi a_0)f_{a_0^+}(m_D^2-m_\pi^2)F_0^{D\pi}(m_{a_0}^2) ~, 
\\
C(a_0 \pi) &=-a_2(a_0 \pi)f_\pi(m_D^2-m_{a_0}^2) F_0^{D a_0}(m_\pi^2) ~,
\\
C'(\pi a_0) &=a_2(\pi a_0)f_{a_0^+}(m_D^2-m_\pi^2)F_0^{D\pi}(m_{a_0}^2) ~,         
\end{split}
\end{align}
where the flavor operators $a_{1,2}$ have been calculated in Ref.~\cite{Cheng:DtoSP} based on
QCD factorization \cite{BBNS}. For the reader's convenience, the relevant part is shown in Table~\ref{tab:aiSP}.
Form factors for $D\to P,S$ transitions are defined by~\cite{BSW}
\begin{align}
\label{eq:DSm.e.}
\begin{split}
\la P(p')|V_\mu|D(p)\ra &= \left(P_\mu-{m_D^2-m_P^2\over q^2}\,q_ \mu\right)
F_1^{DP}(q^2)+{m_D^2-m_P^2\over q^2}q_\mu\,F_0^{DP}(q^2) ~, 
\\
\la S(p')|A_\mu|D(p)\ra &= -i\Bigg[\left(P_\mu-{m_D^2-m_S^2\over
q^2}\,q_ \mu\right) F_1^{DS}(q^2)   +{m_D^2-m_S^2\over
q^2}q_\mu\,F_0^{DS}(q^2)\Bigg] ~,        
\end{split}
\end{align}
where $P_\mu=(p+p')_\mu$ and $q_\mu=(p-p')_\mu$.  As shown in Ref.~\cite{CCH}, a factor of $(-i)$ is needed in the $D \to S$ transition in order for the $D \to S$ form factors to be positive which can also be checked from heavy quark symmetry consideration.  For hadronic $D\to SP$ decays, the relevant form factors are $F_0^{DS}$ and $F_0^{DP}$.  For the transition form factor $F_0^{D\pi}$ and its $q^2$ dependence, we use the result obtained in the covariant confined quark model (CCQM) \cite{Ivanov:2019nqd}.
The parameters $F_0^{DS}(q^2)$ at $q^2=0$ and their $q^2$ dependence for $D \to S$ transitions calculated in CCQM~\cite{Soni:2020sgn}, light-cone sum rules (LCSR)~\cite{Cheng:2017fkw,Huang:2021owr} and QCD sum rules~\cite{Wu:2022qqx}  are exhibited in Table~\ref{tab:FFDtoS}. 
The vector decay constant of the scalar meson is defined as $\la S(p)|\bar q_2\gamma_\mu q_1|0\ra=f_S p_\mu$. However, the vector decay constant of the neutral $a_0^0$ vanishes.  We thus define the scalar constant $\la a_0^0|\bar q q|0\ra=m_{a_0^0}\bar f_{a_0^0}$~\cite{CCY}.  As shown in Refs.~\cite{Cheng:2006,Cheng:2013}, beyond the factorization approximation, contributions proportional to the scalar decay constant $\bar f_S$ can be produced from vertex and hard spectator-scattering corrections. Hence, for neutral $a_0^0$ we have
\begin{align}
\begin{split}
T'(\pi a_0^0) &= a_1(\pi a_0^0)\bar f_{a_0^0}(m_D^2-m_\pi^2)F_0^{D\pi}(m_{a_0}^2) ~,
\\
C'(\pi a_0^0) &= a_2(\pi a_0^0)\bar f_{a_0^0}(m_D^2-m_\pi^2)F_0^{D\pi}(m_{a_0}^2) ~.        
\end{split}
\end{align}
We shall use the value of $\bar f_{a_0^0}=365\pm20$ MeV obtained in Ref.~\cite{CCY} at $\mu=1$~GeV within the framework of QCD sum rules. The corresponding vector decay constant is $f_{a^\pm}\approx \mp1.3$ MeV.  It is na{\"i}vely expected that, for example,  $|T'|\ll |T|$ and $|C'|\ll |C|$ for charged $a_0$.  However, as shown in Ref.~\cite{Cheng:DtoSP}, a realistic calculation yields $|C'|>|C|$ instead.

\begin{table}[t]
\caption{Numerical values of the flavor operators $a_{1,2}(M_1M_2)$ for $M_1M_2=a_0(980)P$ and $Pa_0(980)$ taken from Ref.~\cite{Cheng:DtoSP} at the scale $\mu=\ov m_c(\ov m_c)=1.3$~GeV }
\label{tab:aiSP}
\begin{center}
\begin{tabular}{ l c c | l r r} \hline \hline
  $$ & ~~$a_0(980)^0\pi$~~ & ~~~$\pi a_0(980)^0$~~~ & ~~$$~~  & ~~$a_0(980)^0 K$~~~~~ & $K a_0(980)^0$~~~~ \\
  \hline
 $a_1$ & ~~~$1.292+0.080i$~~~ & ~~$0.037-0.066i$~~ & ~~$a_1$ & $1.295+0.075i$ & $0.037-0.066i$  \\
 $a_2$ & $-0.527-0.172i$ & $-0.080+0.141i$ &  ~~$a_2$ & $-0.533-0.162i$ & $-0.080+0.141i$ \\
\hline\hline
  $$ & ~~$a_0(980)^\pm\pi$~~ & ~~~$\pi a_0(980)^\pm$~~~ & ~~$$~~  & ~~$a_0(980)^\pm K$~~~~~ & $K a_0(980)^\pm$~~~ \\
  \hline
 $a_1$ & ~~~$1.292+0.080i$~~~ & ~~$\pm(-10.04+20.03i)$~~ & ~~$a_1$ & $1.295+0.075i$ & ~~~$\pm(-10.04+20.03i)$  \\
 $a_2$ & $-0.527-0.172i$ & ~~$\pm(23.89-43.14i)$ &  ~~$a_2$ & $-0.533-0.162i$ & ~~~$\pm(23.89-43.14i)$ \\
\hline \hline
\end{tabular}
\end{center}
\end{table}

\begin{table}[t]
\caption{Form factor $F_0^{Da_0}$ at $q^2=0$ in various models. }
 \label{tab:FFDtoS}
  \medskip
\begin{ruledtabular}
\begin{tabular}{l c c c c  }
Transition  & CCQM~\cite{Soni:2020sgn} & LCSR~(I)~\cite{Cheng:2017fkw}
& LCSR~(II)~\cite{Huang:2021owr} & QCDSR~\cite{Wu:2022qqx}   \\
\hline
$D\to a_0(980)$  & $0.55\pm0.02$   & $0.88\pm0.13$    &  $0.85^{+0.10}_{-0.11}$
& $1.058^{+0.068}_{-0.035}$
\\ 
\end{tabular}
\end{ruledtabular}
\end{table}

We see from Table~\ref{tab:DSP} that in the na{\"i}ve $q\bar q$ scheme, the ratios defined in Eq.~(\ref{eq:r}) have the expressions
\be
r_{+/-}=\left| {T'+E \over T+E'}\right|^2, \qquad
r_{+/0}=\left| {T'+C-A+A' \over T+C'+A-A'}\right|^2.
\label{eq:r-ratio-diquark}
\en 
Assuming the negligible weak annihilation, we will have $r_{+/-}=|T'/T|^2$ and $r_{+/0}=|(T'+C)/(T+C')|^2$.  Since $|T'|\ll|T|$, it is expected that both $r_{+/-}$ and $r_{+/0}$ are much less than unity. Indeed, from Table~\ref{tab:BF3body} we obtain $r_{+/-}=0.033$ and $r_{+/0}=0.16$, both too small compared to the measured values given in Eq.~(\ref{eq:r}).  It is also clear from the same table that the absolute branching fractions of $D^+\to a_0(980)^+\pi^0, ~a_0(980)^+\ov{K}^0$ and $D^0\to a_0(980)^+\pi^-, ~a_0(980)^+ K^-$ followed by the strong decays $a_0^+\to \pi^+\eta$ or $a_0^+\to K^+\ov{K}^0$ are all too small compared to the experiment. On the contrary, the predicted branching fraction of $D^0\to a_0(980)^-\pi^+$ followed by $a_0(980)^-\to \pi^-\eta$ is too large compared to the data.

In the 2-quark scheme,  one may argue that the $W$-exchange contributions $E$ and $E'$ could be large enough to render a larger rate for $D^0\to a_0(980)^+\pi^-$ and a smaller one for $D^0\to a_0(980)^-\pi^+$. It was claimed in Ref.~\cite{BESIII:Dtoa0pi} that $|E'/T|>1$ and $|E/E'|^2 \sim 0.3$ (see footnote [32] of Ref.~\cite{BESIII:Dtoa0pi} ) would accommodate the data of $r_{+/-}$. This requires a very large $W$-exchange contribution that is difficult to comprehend.  
For the topologies $E$ and $E'$, their factorizable short-distance contributions are presumably very small owing to helicity suppression. Hence, they are dominated by long-distance effects.
In principle, they can be extracted
from a global fit provided that the relevant data are sufficiently available. Naively, they are expected to have similar sizes with some phase difference between them.

In the tetraquark scheme, we have instead the ratios
\be
r_{+/-}=\left| {T'-\tilde{T}+E \over T+E'}\right|^2, \qquad
r_{+/0}=\left| {T'+C-\tilde{T}+\ov T-A+A' \over T+C'+\tilde{T}-\ov T+A-A'}\right|^2,
\en 
where use of the excellent approximation of $\lambda_s+\lambda_d\approx 0$ has been made, and $\tilde{T}$ and $\ov T$ are the new topological diagrams compared to Eq.~\eqref{eq:r-ratio-diquark}. Unlike the primed amplitudes $T'$ and $C'$ which are greatly suppressed as their factorizable contributions are proportional to the decay constant of the charged $a_0(980)$ meson which is of order 1.3~MeV, the nonfactorizable $\tilde{T}$ and $\ov T$ are not suppressed relative to $T'$ and $C'$.  Keeping only the dominant amplitudes $T$ and $\tilde{T}$, the authors of Ref.~\cite{Achasov:2024nrh} advocate that the ratios $r_{+/-}$ and $r_{+/0}$ lead to a correlation
\footnote{In terms of the notations $a$, $b$ and $e$ in Ref.~\cite{Achasov:2024nrh}, $a$ corresponds to our $\tilde{T}$, $b$ to $T$ and $e$ to $E$. }
\be
r_{+/-}=\left| {\tilde{T}\over T}\right|^2
~, \qquad
r_{+/0}=\left| {\tilde{T} \over T+\tilde{T}}\right|^2 
\quad \Rightarrow \qquad 
r_{+/0}={r_{+/-}\over |\sqrt{r_{+/-}}-1|^2}
~,
\en 
which is indeed respected by the data, provided that both $\tilde T$ and $T$ are real and opposite in sign.  Roughly speaking, $\tilde T/T\approx -2.7$. In reality, the $W$-exchange amplitudes cannot be neglected in the decay $D^0\to a_0^-\pi^+$; otherwise, the predicted rate of $D^0\to a_0^-\pi^+\to \eta \pi^-\pi^+$ will be too large by one order of magnitude. Hence, $\tilde T/T$ should have a large deviation from the above expectation.

\begin{table}[t!]
\caption{Branching fractions of various $D\to a_0(980)P\to P_1P_2P$ decays calculated in the quark-antiquark model and the tetraquark model of the scalar meson $a_0(980)$. Experimental data are taken from Refs.~\cite{PDG,BESIII:Dtoa0pi,BESIII:DtoKSa0}. For simplicity, we have dropped the mass identification for  $a_0(980)$.
}
 \label{tab:BF3body}
  \medskip
\begin{ruledtabular}
\begin{tabular}{l c c c}
$D\to SP; S\to P_1P_2$ &  $q\bar q$ model  & Tetraquark model & Data \\
 \hline
 $D^+\to a_0^+\pi^0; a_0^+\to \pi^+\eta$ & $2.9\times 10^{-4}$ & $(0.7-1.0)\times 10^{-3}$ & $(0.95\pm0.13)\times 10^{-3}$ \\
$D^0\to a_0^+\pi^-; a_0^+\to \pi^+\eta$ & $3.3\times 10^{-5}$ & $(2.6-3.2)\times 10^{-4}$ & $(0.55\pm0.09)\times 10^{-3}$ \\
$D^0\to a_0^+\pi^-; a_0^+\to K^+\ov{K}^0$ & $1.3\times 10^{-5}$ & $(0.8-1.0)\times 10^{-4}$ & $(1.2\pm0.8)\times 10^{-3}$ \\
$D^+\to a_0^+\ov {K}^0; a_0^+\to \pi^+ \eta$ & $1.3\times 10^{-2}$ & $2.7\times 10^{-2}$ & $(2.66\pm 0.13)\%$ \\
$D^0\to a_0^+ K^-; a_0^+\to K^+ \ov{K}^0$ & $2.2\times 10^{-4}$ & $(1.8-2.1)\times 10^{-3}$ & $(1.18\pm 0.36)\times 10^{-3}$ \\
$D^0\to a_0^+ K^-; a_0^+\to \pi^+ \eta$ & $6.5\times 10^{-4}$ & $(5.9-7.0)\times 10^{-3}$ & $(7.4^{+0.9}_{-0.7})\times 10^{-3}$ \\
$D^0\to a_0^0\ov K^0; a_0^0\to \eta\pi^0$ & $1.1\times 10^{-3}$ & $(1.9-3.0)\times 10^{-2}$ & $(2.40\pm0.56)\%$  \\
$D^0\to a_0^0\ov K^0; a_0^0\to K^+K^-$ & $3.4\times 10^{-4}$  & $(5.9-9.2)\times 10^{-3}$ & $(5.8\pm0.8)\times 10^{-3}$    \\
$D_s^+\to a_0(980)^+\pi^0+a_0(980)^0\pi^+\to \pi^+\pi^0\eta$ & $-$  & $(1.6-3.0)\times 10^{-2}$ \footnotemark[1] & $(1.46\pm0.27)\%$   \footnotemark[2] \\
\hline
$D^+\to a_0^0\pi^+; a_0^0\to \pi^0\eta$ & $1.8\times 10^{-3}$ & $(0.9-2.6)\times 10^{-4}$ & $(3.7\pm1.1)\times 10^{-4}$ \\
$D^0\to a_0^-\pi^+; a_0^-\to \pi^-\eta$ & $9.3\times 10^{-4}$ & $0.7\times 10^{-4}$ & $(0.7\pm 0.2)\times 10^{-4}$ \\
$D^0\to a_0^-\pi^+; a_0^-\to K^- K^0$ & $2.9\times 10^{-4}$ & $2.2\times 10^{-5}$ & $(2.6\pm 2.8)\times 10^{-4}$ \\
$D^0\to a_0^-K^+; a_0^-\to K^-\ov K^0$ & $1.7\times 10^{-5}$ & $1.3\times 10^{-6}$ & $<2.2\times 10^{-4}$   \\
\end{tabular}
\footnotetext[1]{$W$-annihilation contributions have been neglected in the calculations.}
\footnotetext[2]{ The branching fraction is assigned to be $(2.2\pm0.4)\%$ by the PDG \cite{PDG}.
However, as pointed out by BESIII \cite{BESIII:Dstoa0pi}, the fraction of $D_s^+\to a_0(980)^{+(0)}\pi^{0(+)}, a_0(980)^{+(0)}\to \pi^{0(+)}\eta$ with respect to the total fraction of $D_s^+\to a_0(980)\pi,a_0(980)\to\pi\eta$ is evaluated to be 0.66. Consequently,
the branching fraction should be multiplied by a factor of 0.66 to become $(1.46\pm0.27)\%$.}
\end{ruledtabular}
\end{table}

In fact, the tetraquark contribution from $\tilde{T}$ can be cleanly inferred from the recent observation of $D^+\to K_S^0 a_0(980)^+$ in the amplitude analysis of $D^+\to K_S^0\pi^+\eta$ by BESIII~\cite{BESIII:DtoKSa0}. This mode has the advantage that it is free of weak annihilation contributions:
\be
A(D^+\to a_0(980)^+\ov{K}^0)=V_{cs}^* V_{ud}(T'+C+\tilde{T})
~. 
\en
We see from Table~\ref{tab:BF3body} that the predicted rate of $D^+\to a_0(980)^+\ov {K}^0\to \pi^+\eta\ov K^0$ in the $q\bar q$ model of $a_0(980)$ is too small by a factor of 2. This discrepancy can be alleviated in the presence of $\tilde{T}$.  It is found that the measured result of $\B(D^+\to a_0(980)^+\ov {K}^0; a_0(980)^+\to \pi^+ \eta)$ can be accommodated with $\tilde{T}/T\sim -0.25$ or $1.2$\,, assuming both $T$ and $\tilde{T}$ are real. We shall see shortly that the second possibility of $\tilde{T}/T=1.2$ is ruled out by the observation of $D^+\to a_0(980)^0\pi^+$.

In the absence of $W$-exchange contributions, the predicted rate of $D^0\to a_0(980)^-\pi^+\to \pi^+\pi^-\eta$ will be too large by one order of magnitude compared to the data (see Table~\ref{tab:BF3body}). Therefore, we need a sizable destructive contribution from $E'$.  Let $E'/T=|E'/T|e^{i\phi}$, where $\phi$ is the relative phase between $E'$ and $T$. We find that $\phi$ lies  between $180^\circ$ and $164^\circ$ and hence $E'/T$ is between $0.725e^{i180^\circ}$ and $0.961e^{i164^\circ}$.

We next turn to $D^0\to a_0^0\ov K^0\to \eta\pi^0\ov K^0$ to extract $\ov T$ as the amplitude is proportional to $(C-E+\ov T)$.  It follows that $\ov T/T=(1.00^{+0.15}_{-0.17})T$ provided that $E\approx E'$ and $\ov T$ is real. Consequently,  the magnitude of $\ov T$ is substantially larger than that of $\tilde T$.

For $D^+\to a_0(980)\pi$ decays, their topological amplitudes are
\begin{align}
\begin{split}
A(D^+\to a_0(980)^+\pi^0)  &= {1\over\sqrt{2}}V_{cd}^*V_{ud}(T'+C-\tilde T+\ov T-A+A')
~, \\
A(D^+\to a_0(980)^0\pi^+)  &= {1\over\sqrt{2}}V_{cd}^*V_{ud}(T+C' +\tilde T-\ov T+A -A')
~.    
\end{split}
\end{align}
We see from Table~\ref{tab:BF3body} that the calculated branching fraction of $D^+\to a_0(980)^0\pi^+\to\eta\pi^0\pi^+$ from the contributions of $T+C'$ alone is too large by a factor of $\sim 3$. This implies that a destructive contribution from the combination $(\tilde T-\ov T)$ is needed.  Hence, this favors the solution of $\tilde T=-0.25T$ and rules out the other one with $\tilde T=1.2 T$.

\subsection{Discussions}

To summarize, using the measurements of $D^+\to a_0^+(\to \pi^+\eta)\ov K^0$, $D^0\to a_0^-(\to \pi^-\eta)\pi^+$ and $D^0\to a_0^0(\to \pi^0\eta)\ov K^0$ as the input, we obtain
\be 
\tilde{T}=-(0.22^{+0.03}_{-0.02})T, \qquad \ov T=(1.00^{+0.15}_{-0.17})T, \qquad E'/T=(0.725e^{i180^\circ}\sim 0.961e^{i164^\circ}). 
\en
The calculated branching fractions of various $D\to a_0(980)P\to P_1P_2P$ decays in the quark-antiquark and tetraquark models of $a_0(980)$ are shown in Table~\ref{tab:BF3body} where we have used the value of LCSR~(II) for the form factor $F_0^{Da_0}$ (see Table~\ref{tab:FFDtoS}). We see that the calculated results are generally in good agreement with experiment except for $D^0\to a_0^+(\to K^+\ov K^0)\pi^-$ and $D^0\to a_0^-(\to K^-K^0)\pi^+$ where the predicted branching fractions are too small by one order of magnitude. However, the current experimental measurements of these two modes have very large uncertainties and need to be improved in the future. 

Before proceeding, we would like to make two remarks about the calculations in Table~\ref{tab:BF3body}. First, 
we have made an assumption that the form factor $F_0^{Da_0}$ and the decay constant $f_{a_0}$ or $\bar f_{a_0}$
are the same in both 2-quark and 4-quark models. This is mainly because the calculations in the tetraquark model are still not available except for some decay constant and the mass of the scalar meson evaluated using QCD sum rules (see the last two papers in Ref.~\cite{Achasov:2021dvt}).  This is certainly a topic that deserves more attention in the theory community. On the experimental front, BESIII will be able to determine the $D\to a_0(980)$ transition form factor in the near future.  Recall that the $D \to f_0(500)$ form factor has already been measured by BESIII recently~\cite{BESIII:2024lnh}.  The practitioners will claim that their favored model, either quark-antiquark or tetraquark, can reproduce the form-factor measurement. Therefore, it is reasonable to assume that form factors or decay constants are similar in the 2-quark and 4-quark models. 
Nevertheless, the predictions in both models can be very different owing to the presence of new $T$-like contributions, as we have advocated in this work. Second, we have neglected the effects of $W$-exchange and $W$-annihilation in the $q\bar q$ model calculations. 

As discussed in Section~\ref{sec:intro}, if $a_0(980)$ is a $q\bar q$ state,  the $D_s^+\to a_0^+\pi^0+a_0^0\pi^+$ decay proceeds only through the $W$-annihilation topology.  However, its branching fraction observed at a percent level is much larger than the other two $W$-annihilation channels $D_s^+\to \omega\pi^+$ and $\rho^0\pi^+$.  Hence, the question is why the annihilation amplitude is so sizable in the $SP$ sector with $|A/T|_{SP}\sim 1/2$~\cite{Cheng:DtoSP}, whereas it is rather suppressed in the  $P\!P$ and $V\!P$ sector with $|A/T|_{PP}\sim 0.18$ and $|A_V/T_P|_{VP}\sim 0.07$~\cite{Cheng:2024hdo}.  Although potentially large long-distance contributions to the $W$-annihilation topology through the triangle diagrams at the hadron level had been evaluated in the literature~\cite{Hsiao:a0,Ling:a0}, the aforementioned question was not really addressed. If $a_0(980)$ is a tetraquark state,  the decay $D_s^+\to a_0^+\pi^0+a_0^0\pi^+$ receives an additional contribution of $\ov T$. This contribution alone is slightly larger than the observed branching fraction (see Table~\ref{tab:BF3body}) and the destructive contribution from $W$-annihilation will help explain the data.  We thus conclude that $D_s^+\to a_0^+\pi^0+a_0^0\pi^+$ is not a purely $W$-annihilation process and that the $W$-annihilation process in the $SP$ sector is not greatly enhanced relative to the $PP$ or $V\!P$ sector.

\section{Conclusions}
\label{sec:Conclusions}

In this work, we have shown that the hadronic decays $D\to a_0(980)\pi$ serve a remarkable place for discriminating the internal structure of the $a_0(980)$ meson. If $a_0(980)$ is a $q\bar q$ state, the predicted branching fractions of $D\to a_0(980)^+\pi$ are too small by one to two orders of magnitude compared to the experiment as the corresponding amplitude is suppressed by the smallness of the $a_0^+$ decay constant, while the predicted rates for $D^+\to a_0(980)^0 P$ and $D^0\to a_0(980)^{-,0}P$ are usually too large compared to the data.  In particular, the ratios $r_{+/-}$ and $r_{+/0}$ na{\"i}vely expected to be at a few percent level are measured to be around 7.5 and 2.6, respectively.

The above difficulties can be resolved in the tetraquark model of $a_0(980)$ as there exist additional $T$-like topological diagrams, namely, $\tilde{T}$ and $\ov T$.  Although both amplitudes are nonfactorizable and thus not calculable, their information can be extracted from data.  We find $\ov T\approx T$ and $\tilde{T}\approx -0.25T$.  A product of this study shows that the decay $D_s^+\to a_0^+\pi^0+a_0^0\pi^+$ is not a purely $W$-annihilation process and it is governed by the tetraquark topology $\ov T$.  Therefore, measurements of $D\to a_0(980)P$ decays lend strong support to the tetraquark picture of $a_0(980)$.

\section*{Acknowledgments}

This research was supported in part by the National Science and Technology Council of R.O.C. under Grant Nos.~112-2112-M-001-026 and 111-2112-M-002-018-MY3 and the National Natural Science Foundation of China under Grant Nos.~12475095 and U1932104. FRX would like to acknowledge the hospitality of Institute of Physics, Academia Sinica, where part of this work was done.

\appendix

\section{Three-body decays of $D\to a_0(980)P\to P_1P_2 P$}

We take the $D^+\to a_0^+\ov K^0\to \pi^+\eta \ov K^0$ decay as an explicit example to illustrate the calculation of the three-body decay rate.  The two-body decay amplitude for $D^+\to a_0^+(m_{12})\ov K^0$ with $m_{12}$ ($s_{12}=m_{12}^2\equiv (p_1+p_2)^2)$ being the mass of $a_0$ is given by
\begin{align}
\begin{split}
A(D^+\to a_0^+(m_{12})\ov K^0)
=&
{G_F\over\sqrt{2}}V_{cs}^*V_{ud}\Big[ a_1(\ov K a_0)f_{a_0^+}(m_D^2-s_{12})F_0^{DK}(s_{12})
\\
&~~~ -a_2(a_0 \ov K)f_K(m_D^2-m_{a_0}^2)F_0^{Da_0}(m_K^2)\Big].
\end{split}
\end{align}
Denoting $\A_{a_0}\equiv A(D^+\to a_0^+\ov K^0\to \pi^+(p_1)\eta(p_2) \ov K^0(p_3))$, we have
\be
\A_{a_0}= g^{a_0\to \eta\pi} F(s_{12},m_{a_0})\,T^{\rm Flatte}_{a_0}(s_{12})A(D^+\to a_0^+(s_{12}) \ov K^0),
\en
where the form factor $F(s,m_{a_0})$ parametrized as $F(s,m_{a_0})=[(\Lambda^2+m_{a_0}^2)/( \Lambda^2+s)]^n$ is introduced to account for the off-shell effect of the scalar resonance $a_0(980)$.  Since $a_0(980)$ couples strongly to the $K\ov K$ and $\pi\eta$ channels, its strong interactions can be described by a coupled formula, the so-called Flatt\'e line shape~\cite{Flatte:1976xu}. Explicitly, 
\be
T^{\rm Flatte}_{a_0}(s)={1 \over s-m_{a_0}^2+i\left[g_{a_0\to\eta\pi}^2\rho_{\eta\pi}(s)+g^2_{a_0\to K\bar K}\rho_{K\bar K}(s)\right]}.
\en
with
\be
\rho_{\eta\pi}(s) 
&=& 
{1\over 16\pi}\left(1-{(m_\eta-m_\pi)^2\over s}\right)^{1/2} \left(1-{(m_\eta+m_\pi)^2\over s}\right)^{1/2},
\en
and 
\be
\rho_{K\!\bar K}(s) 
&=& 
{1\over 16\pi}
\left[
\left(1-{4m_{K^+}^2\over s}\right)^{1/2} + \left(1-{4m_{K^0}^2\over s}\right)^{1/2}
\right],
 \en
for the neutral $a_0(980)^0$ as well as
\be
\rho_{K\!\bar K}(s) &=& {1\over 16\pi}\left(1-{(m_{K^+}-m_{K^0})^2\over s}\right)^{1/2} \left(1-{(m_{K^+}+m_{K^0})^2\over s}\right)^{1/2}, 
 \en
for the charged $a_0(980)^\pm$.  The strong couplings $g_{a_0\to\eta\pi}$ and $g_{a_0\to K\bar K}$ can be found in Ref.~\cite{Cheng:DtoSP}.  The decay rate then reads
\begin{align}
\begin{split}
&
\Gamma(D^+\to a_0^+\ov K^0\to \pi^+\eta \ov K^0)
\\
&= {1\over(2\pi)^3 32 m_D^3}\int ds_{12}\,ds_{23} 
\left|g^{a_0\to \pi\eta} F(s_{12},m_{a_0})T^{\rm Flatte}_{a_0}(s_{12}) A(D^+\to a_0^+(m_{12})\ov K^0) \right|^2.
\end{split}
\end{align}


\end{document}